\documentclass[floatfix,prb,aps,superscriptaddress,showpacs,11pt]{revtex4}
\usepackage{amsmath}
\usepackage{graphicx}

\begin{document}
\title{Effective Optical Response of  Metamaterials}
 \author{Guillermo P.  Ortiz\email[email:]{gortiz@exa.unne.edu.ar}}
 \affiliation{Departamento de
  F\'isica, Facultad de Ciencias Exactas, Universidad Nacional del Nordeste,\\
  Av. Libertad 5400 Campus-UNNE, W3404AAS Corrientes, Argentina.  }
\author{Brenda E.  Mart\'inez-Z\'erega}
 \affiliation{Centro Universitario de los Lagos, Universidad de  Guadalajara,
Enrique D\'iaz de Le\'on SN, Paseos de la Monta\~na, Lagos de Moreno,
Jalisco, C.P. 47460, M\'exico.}
 \affiliation{Division of Photonics,
  Centro de Investigaciones en Optica, \\Le\'on, Guanajuato, M\'exico}
 \author{Bernardo S. Mendoza\email[email:]{bms@cio.mx}}
 \affiliation{Division of Photonics,
  Centro de Investigaciones en Optica, \\Le\'on, Guanajuato, M\'exico}
 \author{W. Luis Moch\'an\email[email:]{mochan@fis.unam.mx}}
 \affiliation{Instituto de Ciencias F{\'\i}sicas, Universidad Nacional
   Aut\'onoma de M\'exico,\\Apdo. Postal 48-3, 62251 Cuernavaca,
   Morelos, M\'exico}
\begin{abstract}
  We use a homogenization
  procedure for 
  Maxwell's equations in order to obtain in the local limit the
  frequency  dependent 
   macroscopic dielectric response tensor $\epsilon^M_{ij}(\omega)$ of
   metamaterials made of a matrix with inclusions of any geometrical
   shape repeated periodically with any lattice structure.
   We illustrate the formalism calculating $\epsilon^M_{ij}(\omega)$  
for   several structures. For dielectric rectangular inclusions within
a conducting material we obtain an anisotropic response which
may change from conductor-like at low $\omega$ to
dielectric-like with resonances at large $\omega$, attaining a very
small reflectance  
at intermediate frequencies 
which can be tuned through geometrical tailoring. A simple explanation
allowed us to predict and confirm 
similar behavior for other shapes,  even isotropic, close
to the percolation threshold.
\end{abstract}
\pacs{78.67.Bf, 77.22.Ch, 78.20.Ci, 78.20.Bh}
\maketitle

\section{Introduction}
Metamaterials are typically binary composites of conventional
materials: a matrix with inclusions of
a given shape, arranged in a periodic structure. 
A theoretical model to predict their macroscopic optical properties is
very desirable.
Since the times of Maxwell, Lord Rayleigh 
and Maxwell-Garnet
up to today, many authors have contributed to the calculation of
the bulk
macroscopic response in terms of the dielectric properties of its
constituents (for example, see 
Refs.[\onlinecite{Etopim(1977),Etopim(1993),Etopim(2002)}]) employing
various approaches such as variational theories or completely general
theories.\cite{Hashin(1962)} 
The macroscopic effective response
can be obtained by
defining the 
microscopic response of a
composite, 
averaging the microscopic fields and eliminating the contribution of
the fluctuating fields to the average of the the microscopic
response.\cite{mochanPRB85a} Furthermore, the accuracy of the
computational method may be confirmed by using general theorems such
as Keller's reciprocal theorem.\cite{keller(1963),keller(1964),nevard(1985)} 

Recent
technologies   allow the manufacture of ordered composite
materials with periodic structures.
For instance,
high resolution electron beam lithography and its interferometric
version 
have been used in
order to make particular designs of nano-structured
composites, producing various shapes with 
nanometric sizes.\cite{Noda(2003),Grigorenko(2005)}
Moreover, ion milling techniques are capable of producing high quality
air hole 
periodic and non-periodic two-dimensional (2D) arrays, where the holes
can have different geometrical shapes.\cite{kleinPRL04,gordonPRL04}  
Therefore,  it is possible
to build devices with novel macroscopic optical properties.\cite{Pendry(2000)} 
For example, a negative refractive index has
been predicted and observed\cite{Shalaev(2005)} for a periodic 
composite structure of a dielectric matrix with noble metal
inclusions of trapezoidal shape.\cite{Kildishev(2006)}

These advances in metamaterial design have motivated 
a renewed interest in the
study of their optical properties, although the
study of the optical properties of composites is not new, and
several important schemes have been developed in the past.
For example, the macroscopic responses
of a bidimensional periodic array of infinite cylinders was
calculated in 1959 in terms of the Hertz's
potential for a two-dimensional scattering 
problem.\cite{Khizhnyak(1959)} 
Rayleigh's
extended method was applied in order to predict the optical properties
of 
a disordered array of spheres.\cite{McPhedran(1977)}
The variation of the conductivity with the filling 
fraction of an ordered array of conducting spheres
on an insulating matrix 
has been studied too,\cite{Doyle(1977)} and the multipolar effects due
to the inhomogeneities of the local 
field have been analyzed\cite{Claro(1984)} for dielectric spheres at high
filling fraction, yielding 
criteria for their importance as a function of interparticle  
separation.\cite{Rojas(1986)} 
Furthermore, a general theory was developed to describe the
electromagnetic response without any reference to a
specific representation, resulting on a powerful tool to calculate the
macroscopic dielectric response.\cite{mochanPRB85b}  
For periodic composites, a Fourier 
representation is most fitting and expressions for the bulk
macroscopic response may be written in terms of the Fourier
coefficients of the microscopic response.\cite{liJOSAA97,taoPRB90,Shen(1990),krokhinPRB02,haleviPRL99,dattaPRB93}
On the other hand, a spectral representation theory has allowed 
the separation of geometric from material 
properties,\cite{Fuchs(1977),miltonJAP81,bergmanPRB92}
and it has been employed to study 
the transport properties of several
systems.\cite{Etopim(1993),Etopim(2002)}

In connection with nano-structured metallic films there has been some
important development as well.
An exact eigenfunction formulation,\cite{Sheng(1982)} and an approximate modal
formalism,\cite{Lochbihler(1993)} were used to explain resonances in the zeroth
diffraction order of silver square-wave gratings,\cite{Sheng(1982)} and
gold-wire gratings.\cite{Lochbihler(1994)}
 In these works, it was found
that resonances might appear due to the excitation of surface modes.
Such modes can be excited if their momentum matches that of the incident
light after being diffracted by some reciprocal lattice vector of the
periodically structured metal surface. Thus, surface 
plasmon-polariton (SPP) modes
are excited on  the metal-air interface yielding several related
phenomena such as an enhancement of optical
transmission through sub-wavelength holes.\cite{Ebbesen(1998)} 
Beside the single coupling to SPP modes, double
resonant  conditions\cite{Zayats(2003)} and  waveguide
modes\cite{Porto(1999)} seem to play an important role in the enhancement 
for metallic gratings with very narrow slits and for compound 
gratings.\cite{Skigin(2005)} 

A very strong polarization dependence in the optical
response of periodic arrays of oriented sub-wavelength holes on metal
hosts has been recently reported,\cite{kleinPRL04,gordonPRL04} as
well as for a single rectangular inclusion within a perfect
conductor.\cite{Garcia(2005)} The studies above do not rely on SPP
excitation as a mechanism to explain the optical results.

In this work we obtain the macroscopic dielectric response of a
periodic composite,
using a homogenization procedure first proposed by Moch\'an and
Barrera \cite{mochanPRB85a} within the context of the local field
effect at crystals, liquids and disordered composites. In this
procedure the macroscopic response of the system is obtained from its
microscopic constitutive equations by eliminating the spatial
fluctuations of the 
field with the use of Maxwell's equations and solving for the
macroscopic displacement in terms of the macroscopic electric field. 
Besides the average dielectric function, 
the formalism above incorporates the effects that the
rapidly varying Fourier components of the microscopic response has on
the macroscopic response. 
An equivalent
procedure suitable for periodic systems 
was recently proposed by P.\ Halevi and
F.\ P\'erez-Rodr\'{\i}guez\cite{Perez(2006)} and applied to photonic
crystals and  
metamaterials. Although developed independently, it
may be considered an extension of the 
generalized local field effect theory developed previously by Moch\'an
and Barrera\cite{mochanPRB85a} and it has been applied to the
dielectric, magnetic and in general, the bi-anisotropic response of
photonic crystals. 
Similar homogenization procedures are also found
in Refs.~\onlinecite{krokhinPRB02,haleviPRL99,dattaPRB93}.
We further
restrict ourselves to the local limit, in which we neglect the
dependence of the response on the wavevector, or more precisely, on
the Bloch's vector.
The macroscopic
optical response is obtained in  terms of the geometrical shape of the
inclusions, their 
periodic arrangement, and the dielectric function of the host and the
inclusions. The proposed scheme is straightforward, requiring
standard numerical computations. It has the advantage
of fully accounting  for the detailed geometry of the system.
For systems with periods much smaller than the wavelength of the incoming
light, the local limit becomes the exact response while it accounts
for the local 
field effect, i.e., the interaction among parts of the system through
the spatially fluctuating electromagnetic field.
We reproduce,  
previously reported
results,\cite{Milton(1981)b,taoPRB90,bergmanPRB92} 
and novel effects resulting solely from the
geometrical shape of the inclusions, namely, the existence of
transparency windows within metal-dielectric metamaterials slightly
above the percolation threshold of the metallic phase.

The article is organized as follows. In Sec. \ref{gp} we present the
theoretical approach used for the calculation of the macroscopic
dielectric response of the composite. 
In Sec. \ref{results}
we validate our formalism comparing it with previous
schemes, yielding very 
good agreement. Then, we present results for two-dimensional
periodic structure consisting of a gold host with  dielectric
rectangular prism or circular inclusions. 
Finally,
in Sec. \ref{conc} we present our conclusions.

\section{Theoretical Approach}\label{gp}
In order to calculate the macroscopic dielectric response of a
metamaterial we follow the steps of Ref. \onlinecite{mochanPRB85a}.
We start by defining appropriate average and fluctuation idempotent projectors
$\hat P_a$ and $\hat P_f=\hat 1-\hat P_a$ such that $\hat P_a$ acting
on any microscopic field $\mathbf F$ produces its macroscopic
projection $\mathbf F^M\equiv \mathbf F_a \equiv \hat P_a \mathbf F$,
while $\hat P_f$ acting on the same field 
yields the spatially fluctuating part $\mathbf F_f \equiv \hat P_f\mathbf F=\mathbf F-\mathbf
F^M$ which we wish to eliminate. The constitutive equation $\mathbf
D=\hat \epsilon \mathbf E$, where $\hat\epsilon$ is the dielectric
operator (in the general case, a complex tensorial integral operator for each
frequency), may be split
into macroscopic and spatially fluctuating parts. Thus we 
write
\begin{equation}\label{DMvsE}
  \mathbf D^M=\hat\epsilon_{aa}\mathbf E^M + \hat\epsilon_{af}\mathbf E_f, 
\end{equation}
where $\hat O_{\alpha\beta}=\hat P_\alpha \hat O \hat
P_\beta$ ($\alpha,\beta=a,f$) for any operator $\hat O$ and we used
the idempotency of the projectors. Furthermore, the
fluctuating part of the wave equation for a non-magnetic material is
given by
\begin{equation}\label{onda}
  \nabla\times(\nabla\times \mathbf E_f) = k_0^2 \mathbf D_f = k_0^2
  (\hat \epsilon_{fa} 
  \mathbf E^M + \hat \epsilon_{ff} \mathbf E_f),
\end{equation}
where $k_0=\omega/c=2\pi/\lambda_0$ and $\lambda_0$ are the free space
wavenumber and wavelength corresponding to frequency $\omega$ and we
assumed that 
the external sources have no spatial fluctuations (otherwise, a
homogenization procedure would prove useless). We solve
Eq. \eqref{onda} for the fluctuating electric field
\begin{equation}\label{ef}
  \mathbf E_f = - \left(\hat\epsilon_{ff} - \frac{1}{k_0^2}
  (\nabla\times\nabla\times)_{ff} \right)^{-1} \hat \epsilon_{fa} 
  \mathbf E^M 
\end{equation}
where $\nabla\times\nabla\times$ denotes the operator
($\mathrm{grad}\,\mathrm{div} -\nabla^2)$ and  the inverse on the RHS
may be interpreted in real space as a 
Green's function, i.e., an integral
operator whose kernel obeys a differential equation with a singular
source.  The inverse in the
second term in the RHS of 
Eq. (\ref{ef}) is 
performed {\em after} the projections onto the space of fluctuating fields,
denoted by the two subscripts $ff$. Finally,
we substitute Eq. \eqref{ef} into Eq. \eqref{DMvsE} to obtain the
macroscopic relation $\mathbf D^M = \hat \epsilon^M \mathbf E^M$ where
we identify the macroscopic dielectric operator
\begin{equation}\label{mb21}
  \hat\epsilon^M = \hat\epsilon_{aa} - \hat\epsilon_{af}
  \left(\hat\epsilon_{ff} -
  \frac{1}{k_0^2}(\nabla\times\nabla\times)_{ff}\right)^{-1} 
  \hat\epsilon_{fa}.
\end{equation}
The first term in the RHS of
Eq. (\ref{mb21}) represents the average dielectric response, while the
second term incorporates the effect of the interactions through the
small-lengthscale spatial fluctuations of the field on the macroscopic
response. 

We rewrite Eq. (\ref{mb21}), which corresponds to Eq. (21) of
Ref. \onlinecite{mochanPRB85a}, as
\begin{equation}\label{niuno}
  \hat\epsilon^M = \hat\epsilon_{aa} - \hat\epsilon_{af}
  \hat \Phi_{fa},
\end{equation}
where $\hat\Phi_{fa}$ is defined through 
\begin{equation}\label{Phi}
  \hat{\mathcal W}_{ff}\hat \Phi_{fa} = \hat \epsilon_{fa} 
\end{equation}
and we introduced the wave operator
\begin{equation}\label{Wave}
  \hat{\mathcal  W} = \hat \epsilon - \frac{1}{k_0^2} \nabla\times\nabla\times.
\end{equation}

For a periodic system, we can use Bloch's theorem to represent the
fields and operators through their Fourier components
\begin{equation}\label{er}
\mathbf{F}_{\mathbf{q}}(\mathbf{r})=\sum_{\mathbf{G}}\mathbf{F}_\mathbf{q}(\mathbf{G})e^{i(\mathbf{q}+\mathbf{G})\cdot\mathbf{r}}, 
\end{equation}
\begin{equation}\label{Or}
{\cal O}_{\mathbf{q}}(\mathbf{r},\mathbf{r}') = \sum_{\mathbf{G}\mathbf{G}'} {\cal O}_\mathbf{q}(\mathbf{G},\mathbf{G}')
e^{i[(\mathbf{q}+\mathbf{G})\cdot\mathbf{r}-(\mathbf{q}+\mathbf{G}') \cdot\mathbf{r}']}, 
\end{equation}
where $\mathbf{F}_{\mathbf{q}}(\mathbf{r})$  denotes an arbitrary position dependent field with
a given Bloch vector $\mathbf{q}$,
${\cal O}_{\mathbf{q}}(\mathbf{r},\mathbf{r}')$ is the kernel corresponding to an arbitrary
operator $\hat{{\cal O}}$ for the same Bloch vector, and $\mathbf{G}$,
$\mathbf{G}'$ are reciprocal vectors. In this case
we can chose $\hat P_a$ as a truncation operator in reciprocal space
that eliminates the Fourier components outside of the first Brillouin
zone, which can be represented by a Kronecker's delta $\hat P_a \to
\delta_{\mathbf{G}0}$. Also, we can identify $\nabla$ with a diagonal block matrix
$i(\mathbf{q}+\mathbf{G})\delta_{\mathbf{G}\mathbf{G}'}$.
Thus, we rewrite Eqs. \eqref{niuno}-\eqref{Wave}
as  
\begin{equation}\label{nidos}
  \left[\epsilon^M_\mathbf{q}\right]_{ik} = \left[\epsilon_\mathbf{q}(\mathbf
    0,\mathbf0)\right]_{ik} - \sum_j \sum_{\mathbf{G}\ne
    0} \left[\epsilon_\mathbf{q}(\mathbf0,\mathbf{G})\right]_{ij}
   \left[\Phi_\mathbf{q}(\mathbf{G},\mathbf0)\right]_{jk},
\end{equation}
\begin{equation}\label{Phi2}
  \sum_j\sum_{\mathbf{G}'\ne 0} \left[{\mathcal
      W}_\mathbf{q}(\mathbf{G},\mathbf{G}')\right]_{ij} \left[\Phi_\mathbf{q}(\mathbf{G}',\mathbf
  0)\right]_{jk} =  \left[\epsilon_\mathbf{q}(\mathbf{G},\mathbf0)\right]_{ik}, 
\end{equation}
and 
\begin{equation}\label{Wave2}
  \left[{\mathcal  W}_\mathbf{q}(\mathbf{G},\mathbf{G}')\right]_{ij} =
  \left[\epsilon_\mathbf{q}(\mathbf{G},\mathbf{G}')\right]_{ij}  + 
  \frac{1}{k_0^2} \delta_{\mathbf{G}\mathbf{G}'} \sum_{kl}
  \delta_{il}^{kj}(q_k+G_k)(q_l^{\strut}+G_l). 
\end{equation}
As the fields are  vector valued for each reciprocal vector, our
operators are matrix valued for each pair of reciprocal vectors. Thus,
in the equations above we introduced explicitly the Cartesian indices
$ijkl$. We also introduced the usual four-index delta function
$\delta_{il}^{kj}=\delta_{ik}\delta_{lj}-\delta_{ij}\delta_{lk}$. Notice that $\mathbf{G}$
and $\mathbf{G}'$ are different from zero in
Eqs. \eqref{nidos}-\eqref{Wave2}, as they involve the fluctuating
fields. Our Eqs. \eqref{nidos}-\eqref{Wave2} are  closely related to
Eq. (35) of Ref. \onlinecite{Perez(2006)}. We remark that in the long
wavelength limit $G/k_0\gg1$, so that the transverse part of the RHS of
Eq. \eqref{Wave2}  is dominated by its large second term. Thus, from
Eq. \eqref{Phi2}, the
transverse part of $\Phi_{\mathbf q}(\mathbf G,\mathbf 0)$ becomes small, of the
order of $k_0^2/G^2$. Nevertheless, the second term on the RHS of
Eq. \eqref{Wave2} does not affect the longitudinal part of ${\mathcal
  W}_\mathbf{q}(\mathbf{G},\mathbf{G}')$, so that the longitudinal part of
$\Phi_\mathbf{q}(\mathbf{G},\mathbf 0)$ becomes dominant in this limit. This means
that in the 
long wavelength limit, the fluctuations are mostly
longitudinal\cite{mochanPRB85a} and we may neglect retardation in
their calculation.

We consider now a
two-component system made up of a homogeneous host with a local
isotropic dielectric function  $\epsilon_h$, in which arbitrarily shaped
particles with a
local isotropic dielectric function 
$\epsilon_p$  are periodically included.
Then,
\begin{equation}\label{chign}
\left[\epsilon_\mathbf{q}(\mathbf{G},\mathbf{G}')\right]_{ij} = \left[\epsilon_h \delta_{\mathbf{G},\mathbf{G}'} +
\epsilon_{ph}S(\mathbf{G}-\mathbf{G}')\right]  \delta_{ij},
\end{equation}
where $\epsilon_{ph}\equiv\epsilon_p-\epsilon_h$. The Fourier coefficients
\begin{equation}\label{fgo}
S(\mathbf{G})=
\frac{1}{\Omega}\int
S(\mathbf{r}) e^{i\mathbf{r}\cdot\mathbf{G}} d\mathbf{r}
=\frac{1}{\Omega}\int_v
e^{i\mathbf{r}\cdot\mathbf{G}} d\mathbf{r}
,
\end{equation}
characterize completely  the shape of the particle, as the
integrals are over the volume $v$ occupied by the inclusions within a
single unit cell whose total volume is 
$\Omega$. Here, we introduced the characteristic function $S(\mathbf{r})$ whose
value is $S(\mathbf{r})=1$ within $v$ and $S(\mathbf{r})=0$ outside $v$.
In particular,
\begin{equation}\label{fg0}
S(\mathbf{G}=\mathbf0)=  v/\Omega \equiv f
,
\end{equation}
with $f$ the
filling fraction of the inclusions, and
\begin{equation}\label{chig0}
\left[\epsilon_q(\mathbf0,\mathbf0)\right]_{ij}=(\epsilon_h+\epsilon_{ph} f )\delta_{ij}
.
\end{equation}
Notice that for local media $[\epsilon_\mathbf{q}(\mathbf{G},\mathbf{G}')]_{ij}$ depends only
on the difference $\mathbf{G}-\mathbf{G}'$ and it does not depend on $\mathbf{q}$.  

Finally, substituting Eq. \eqref{chign} in Eq. \eqref{nidos} and taking
the local $\mathbf{q}\to\mathbf0$ limit,  we obtain
\begin{equation}\label{aphiM2}
\epsilon^{M}_{ij}\equiv [\epsilon^{M}_{\mathbf0}]_{ij}=
(\epsilon_h + \epsilon_{ph}f)\delta_{ij} 
-  \epsilon_{ph} \sum_{\mathbf{G}\ne 0}
S(-\mathbf{G}) [\Phi_{\mathbf0}(\mathbf{G},\mathbf0)]_{ij}
,
\end{equation}
where $[\Phi_{\mathbf0}(\mathbf{G},\mathbf0)]_{ij}$ is obtained by solving 
Eq. \eqref{Phi2} after substituting Eq. \eqref{chign} and
\begin{equation}\label{W0}
  \left[\mathcal W_{\mathbf 0}(\mathbf{G},\mathbf{G}')\right]_{ij} = 
  \left[\epsilon_h \delta_{\mathbf{G},\mathbf{G}'} + \epsilon_{ph}S(\mathbf{G}-\mathbf{G}')\right]  \delta_{ij}
 - 
  \frac{1}{k_0^2} (G^2 \delta_{ij}-G_i G_j) \delta_{\mathbf{G}\mathbf{G}'}
\end{equation}
from Eq. \eqref{Wave2}. 
Notice that in principle we could take the local limit $\mathbf{q}\to\mathbf
0$ without also taking the long wavelength limit $k_0\to 0$, although
it is advisable to verify that $\epsilon^M_{\mathbf{q}}$ is close to
$\epsilon^M_{\mathbf 0}=\epsilon^M$ for the relevant wavevectors $\mathbf{q}$ that
appear in each particular application.
We remark that the first term on the RHS of Eq. \eqref{aphiM2} is
isotropic as it is simply the average of the response of the
constituents, which we took to be local, piecewise homogeneous and
isotropic. Nevertheless, the second term includes information on the
geometry of the system, including both the shapes of the particles and
their periodic arrangement. Thus, in general it yields a non-isotropic
contribution to the macroscopic dielectric tensor. 

In the following section we show several examples of this
procedure to calculate the macroscopic dielectric tensor
$\epsilon^M_{ij}$.

\section{Results}\label{results}

\subsection{Comparison to Previous Work}\label{previo}

\begin{figure}
\centering
\includegraphics{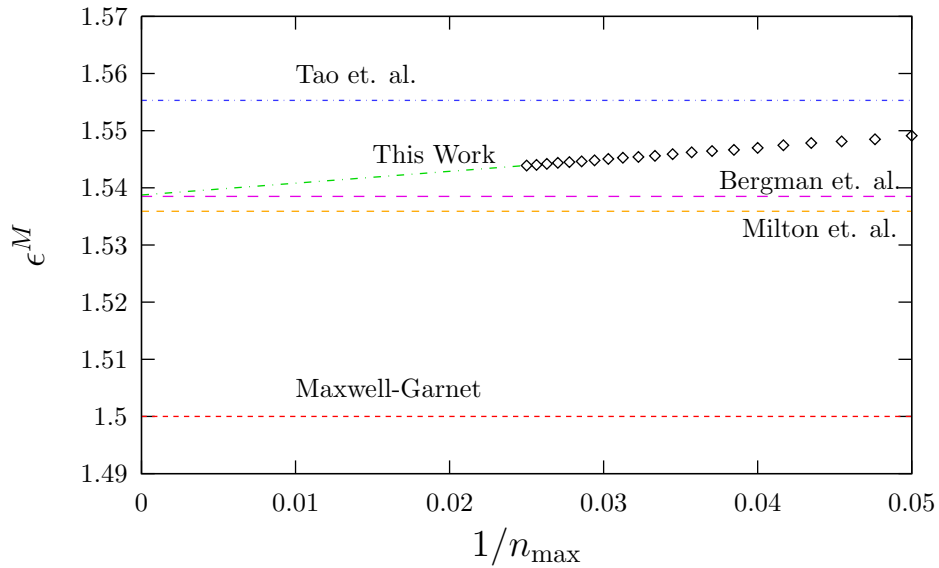}
\caption{(color online)
We show the normal-to-the-axis macroscopic dielectric function $\epsilon^M$
obtained through 
Eq.~\eqref{aphiM2} for a 2D square 
array with lattice parameter $L$ of square dielectric prisms with
response $\epsilon_p=5.0$ placed in vacuum 
with filling fraction $f=0.3$ for a free-space wavelength
$\lambda_0=10L$ as a function of the largest reciprocal vector
index vs. $1/n_{\mathrm{max}}$. 
We show a linear extrapolation of our results toward
$1/n_{\mathrm{max}}\to0$ and we indicate the values predicted by 
Maxwell-Garnet formula and by some other authors mentioned in
the  text.
}
\label{fig:npw}
\end{figure}

In this section we apply our results to light moving across a 2D
square array of infinite square dielectric prisms with diagonals
aligned with the sides of the square primitive cell, a system
previously proposed by  
Milton et
al.\cite{Milton(1981)b}
We chose the parameters
$\epsilon_p=5.0$,
$\epsilon_h=1.0$, and $f=0.3$. We take 
a finite free-space wavelength $\lambda_0=10L$, with $L$ the lattice parameter, so
that, according to Eq. \eqref{W0} we expect only small retardation
effects of the order of $(L/\lambda_0)^2=1/100$.  
We choose the polarization normal to the prisms axis so that in our
local limit the system is effectively isotropic in 2D.
We truncated our matrices in reciprocal space by setting a maximum
value $2\pi n_{\mathrm{max}}/L$ for the magnitude $|G_x|$ and $|G_y|$ of
the components of the reciprocal vectors, so for a field polarization
within the plane 
the number
of  rows and columns for the matrix $\left[\mathcal W_{\mathbf
    0}(\mathbf{G},\mathbf{G}')\right]_{ij}$  in 
Eq.~\eqref{W0} is given by
$8n_{\mathrm{max}}(n_{\mathrm{max}}+1)$. 
To test the convergence of our computational procedure, in
Fig.~\ref{fig:npw} 
we show our results for $\epsilon^M_{ij}\equiv\epsilon^M\delta_{ij}$ as a function
of the maximum index $n_{\mathrm{max}}$. 
From
Fig.~\ref{fig:npw} we see that $\epsilon^M$ converges approximately as 
$1/n_{\mathrm{max}}$, and
values of the order around  $n_{\mathrm{max}}=40$ are needed to obtain
an accuracy better than 0.5\% without extrapolating, 
yielding large matrices of more  than  $13000\times 13000$
elements. In the same figure we have indicated the response obtained
by Milton et 
al.,\cite{miltonJAP81}, Tao et al.,\cite{taoPRB90} and Bergman et
al.\cite{bergmanPRB92} 
which studied the same composite. As we see,  linear extrapolation of
our results towards $1/n_{\mathrm{max}}\to0$ converge to 
those Bergman et al. and of Milton et al., whereas the 
result of Tao et al. 
differs slightly. 
Finally, we also compare our results with those of mean-field theory,
as embodied in Maxwell-Garnet's (MG) formulae
\begin{equation}\label{MG}
\epsilon^{M}=\epsilon_h+f\epsilon_{ph}
-
\frac{\epsilon_{ph}^2f(1-f)}{\gamma\epsilon_h+\epsilon_{ph}(1-f)},
\end{equation}
with $\gamma=2$ for our 2D system.\cite{dattaPRB93}
As we see in the figure, the MG results differ from ours and the other
authors' results, mainly due to its intrinsic limitations.\cite{dattaPRB93}
We have 
checked our results with other set of parameters also reported by
the same cited authors
and we have obtained similar agreement 
as  mentioned above. The rate of convergence of our method is similar
to that reported in Ref. \onlinecite{Sozuer(1992)} when written in terms of
$n_{\mathrm{max}}$. 

\begin{figure}
\centering
\includegraphics{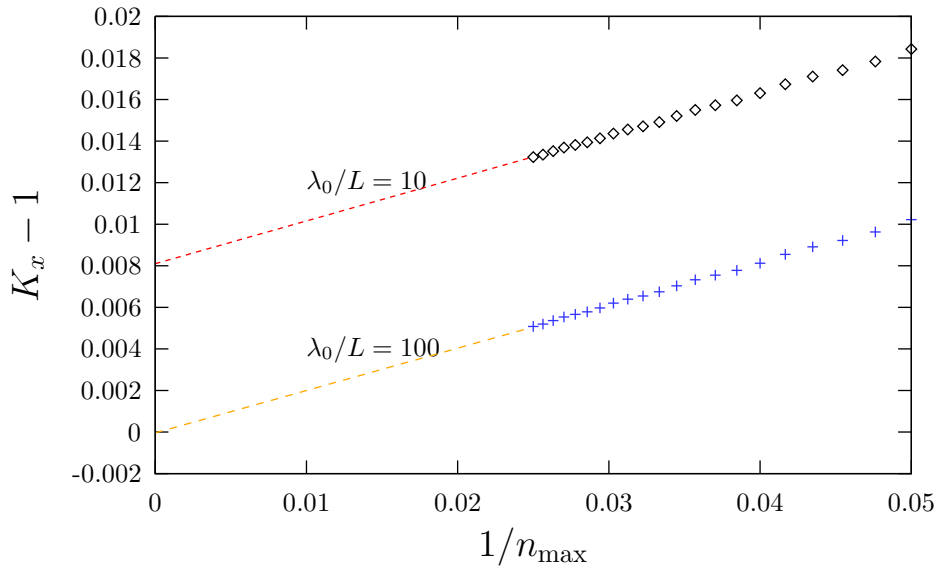}
\caption{(color online) Keller's coefficient $K_x-1$
    as a function of $1/n_{\mathrm{max}}$ and its linear extrapolation 
    towards $n_{\mathrm{max}}\to\infty$ for the same system as in
    Fig. \ref{fig:npw} in the cases of $\lambda_0/L=10$ and 100.} 
\label{fig:keller}
\end{figure}

We can also test the convergence of our results above using Keller's
theorem,\cite{keller(1963),keller(1964),nevard(1985)} which we may
write as  
$K=K_x K_y=1$, where we define Keller's coefficients along principal
axes $x,y$ as $K_x=(\epsilon^M_x\tilde\epsilon^M_x)/(\epsilon_h\epsilon_p)$ and
$K_y=(\epsilon^M_y\tilde\epsilon^M_y)/(\epsilon_h\epsilon_p)$. Here, we introduced the
macroscopic response $\tilde\epsilon^M_x$ and $\tilde\epsilon^M_y$ of
the reciprocal system that is obtained from the original system by
interchanging 
$\epsilon_h\leftrightarrow\epsilon_p$. Indeed, for our isotropic system
we expect $K_x=1$ as $K_x=K_y$. 
In Fig. \ref{fig:keller} we show $K_x-1$ vs. $1/n_{\mathrm{max}}$ for
the system corresponding to Fig. \ref{fig:npw}.
We see clearly that $K_x-1$ decreases linearly in
$1/n_{\mathrm{max}}$. However, its extrapolation towards
$n_{\mathrm{max}}\to\infty$ does not attain the value $K_x-1=0$ as
expected from Keller's theorem. The reason for the small discrepancy is
that our calculation includes retardation effects which we expect to
be of order $(L/\lambda_0)^2$, while Keller's theorem is strictly
valid only in systems with no retardation. To confirm this statement,
we also display in Fig. \ref{fig:keller} the results of a calculation
for $\lambda_0=100L$, showing that in this case, the discrepancy
between the extrapolated and the expected value is negligible. Thus,
we have verified that our calculation is consistent with Keller's
theorem in
the absence of retardation  and has an error that goes to zero as
$1/n_{\mathrm{max}}$ when 
$\lambda_0/L\to\infty$.

\begin{figure}
\centering
\includegraphics{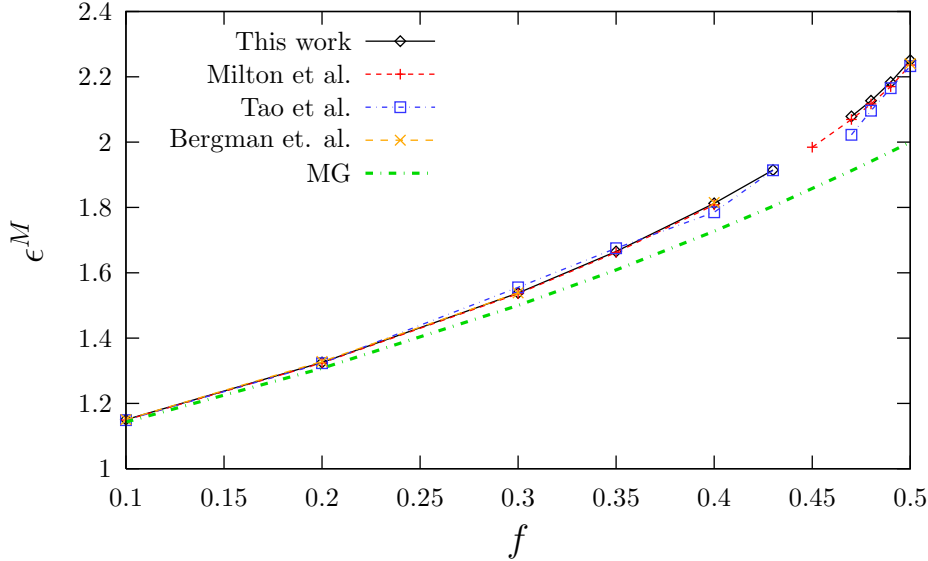}
\caption{(color online) $\epsilon^M$ versus  the filling fraction for the
  same 
  system as shown on Fig.~\ref{fig:npw}. Our result was obtained from
  Eqs.~\eqref{aphiM2} and \eqref{W0} employing a $13120\times13120$
  matrix $\left[\mathcal   W_{\mathbf 0}(\mathbf{G},\mathbf{G}')\right]_{ij}$.
}
\label{fig:comparaFF}
\end{figure}
In Fig.~\ref{fig:comparaFF} we show nearly converged (error $< 0.5\%$)
results for $\epsilon^{M}$
as a function of the filling fraction $f$ for the same system as in
Fig.~\ref{fig:npw}.  We can see again an excellent 
agreement of our results with those obtained by Milton et al., and Bergman
et al.,  and, to a lesser extent, with those of Tao et al.
The MG results are noticeably lower, with a discrepancy that increases
with filling fraction.
We have obtained results identical to ours in Figs. \ref{fig:npw} and
\ref{fig:comparaFF} using Eq. (35) of Ref. \onlinecite{Perez(2006)},
confirming that our formalism is equivalent to that of Halevi and
P\'erez-Rodr\'{\i}guez.  In conclusion, our approach does indeed
reproduce the results 
of other works, and thus we have validated our numerical scheme and
can be confident on the accuracy of our results.

\subsection{2D array}\label{2darray}

Having confirmed our calculation procedure through comparison to
earlier works and convergence tests, we proceed to evaluate the
optical properties of a 
metamaterial.
\begin{figure}[t]
\centering
\includegraphics{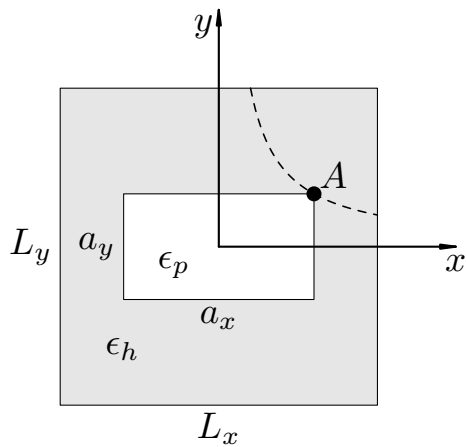}
\caption{
  Unit cell of a 2D
  rectangular array of rectangular prisms with response
  $\epsilon_p$ within a host with response $\epsilon_h$. 
The aspect ratio of the rectangles is determined by the point $A=(L_x \xi_x,  L_y f/\xi_x)$ that lies
on the dashed line from $(L_x f,L_y)$ to $(L_x,L_y f)$. 
}
\label{system}
\end{figure}
We choose a 2D rectangular lattice of rectangular prisms, assuming
translational symmetry along $z$ (see Fig. \ref{system}). 
The unit cell has lengths 
$L_x$ and $L_y$
along the $x$ and $y$ directions, and the inclusions have
corresponding sizes $a_x$ and $a_y$. The shape of the lattice is
controlled by a parameter $\eta$ defined through
\begin{equation}\label{lxvsly}
L_x=\eta L_y
,
\end{equation}
and we define
\begin{equation}\label{zetax}
\xi_i\equiv\frac{a_i}{L_i}\quad i=x,y
.
\end{equation}
Then, 
\begin{equation}\label{gvec}
\mathbf{G}=n_x\frac{2\pi}{L_x}\hat x
+
n_y\frac{2\pi}{L_y}\hat y
=
\frac{2\pi}{L_x}(n_x\hat x
+
\eta n_y\hat y)
,
\end{equation}
for integer $n_x$ and $n_y$, 
and
\begin{equation}\label{ffdef}
f=\xi_x\xi_y
.
\end{equation}
From Eq.~\eqref{fgo} we obtain
\begin{equation}\label{fgrect}
S(\mathbf{G})=
\mbox{sinc}(\frac{G_xa_x}{2})
\mbox{sinc}(\frac{G_ya_y}{2})
=\mbox{sinc}(\pi \xi_xn_x)
\mbox{sinc}(\pi\frac{f}{\xi_x}n_y)
,
\end{equation}
where $\mbox{sinc}(x)=\sin(x)/x$. 
We can vary the shape of the inclusion keeping the filling
fraction fixed by simply changing $\xi_x$ within the bounds
\begin{equation}\label{varxi}
f\le \xi_x \le 1
.
\end{equation}
The array is square if $L_x=L_y$. Furthermore, if $\xi_x=\xi_y=\sqrt{f}$ the
inclusions have a square cross section, while for $\xi_x>\sqrt{f}$ 
($\xi_x<\sqrt{f}$) they become elongated along $x$ ($y$).
For $\xi_x=1$, $\xi_y=f$ ($\xi_x=f$, $\xi_y=1$) the inclusions  fully occupy the  unit
cell along $x$ ($y$), contacting neighbor inclusions, so that the
systems becomes an effectively one dimensional system of slabs with
surfaces normal to $y$ ($x$).

To interpret the results easily we consider a semi-infinite slab $z>0$
cut out of our
metamaterial and we calculate its normal incidence reflectance
\begin{equation}\label{rn}
R_\zeta = \left| \frac{\sqrt{\epsilon^{M}_{\zeta\zeta}}-1}
{\sqrt{\epsilon^{M}_{\zeta\zeta}} + 1} \right|^2 
,\quad(\zeta=x,y)
\end{equation}
corresponding to a $\zeta=x,y$ linearly polarized
incoming beam propagating through empty space along $z$ and impinging upon
the interface which we locate at $z=0$. In this equation we have
neglected the possibility of a magnetic permeability $\mu\ne 1$, which
may be expected even when the constituents of the system are
non-magnetic due to the possible non-locality of the macroscopic
dielectric response 
$\epsilon^M_{\mathbf{q}}$, as may be obtained from Eq. \eqref{nidos}. The
non-locality may be
partially accounted for by a local dielectric function
$\epsilon^M=\epsilon^M_{\mathbf 0}$ and a local magnetic permeability $\mu$
which is of the order of
$\mu-1\sim(\epsilon^M_{\mathbf{q}}-\epsilon^M_{\mathbf 0})/q^2$.\cite{Perez(2006)}
 From Eq. \eqref{Wave2}, we
expect $\mu-1\sim k_0^2 L^2$ where $L$ is of the order of the
periodicity of the system. Another criteria that has been developed for
conducting structures states that the magnetic response may be
neglected as long as the cross section of the particles is much
smaller than the penetration depth.\cite{Krokhin(2007)} 
Thus, in the examples that follow we may safely neglect the magnetic
permeability.

In the following figures we choose
a square unit cell with $L_x=L_y=40$ nm with gold in the interstitial region, 
for which we use the experimentally determined response 
$\epsilon_h=\epsilon\mbox{(Au)}$,\cite{Palik(1985)}
and with dielectric inclusions for which we chose  $\epsilon_p=4$. 
For different values for the filling
fraction $f$ we 
control  the rectangular geometry 
 of the inclusion with the parameter $\xi_x$.
The value of $n_{\mathrm{max}}$ is set to 50
which gives good 
converged results.\footnote{The numerical burden of such large
  matrices has to be surmounted with the use of ScaLAPACK subroutines
(http://www.netlib.org/scalapack/)
  to efficiently solve Eq.~\eqref{Phi2} on a high-end computer
  cluster. Typical time on 40-processor grid is 1.3 hours per energy point.}  

We start with the extreme case $\xi_x=1$, for which we have a system
of alternating conductor and dielectric flat slabs piled up along the
$y$ direction. In this case, the non-retarded macroscopic dielectric
response is given exactly by 
\begin{equation}\label{exM}
\epsilon^{M}_{xx}=f\epsilon_{p}+(1-f)\epsilon_h
,
\end{equation}
and
\begin{equation}\label{eyM}
\frac{1}{\epsilon^{M}_{yy}}=\frac{f}{\epsilon_p}+\frac{1-f}{\epsilon_h}
.
\end{equation}
The latter expression can be written as 
\begin{equation}\label{eyMG}
\epsilon^{M}_{yy}=\epsilon_h+f\epsilon_{ph}
-
\frac{\epsilon_{ph}^2f(1-f)}{\epsilon_h+\epsilon_{ph}(1-f)}
,
\end{equation}
which is the MG result for one dimension, i.e., Eq.~\eqref{MG} with
$\gamma=1$. 
\begin{figure}
\centering
\includegraphics{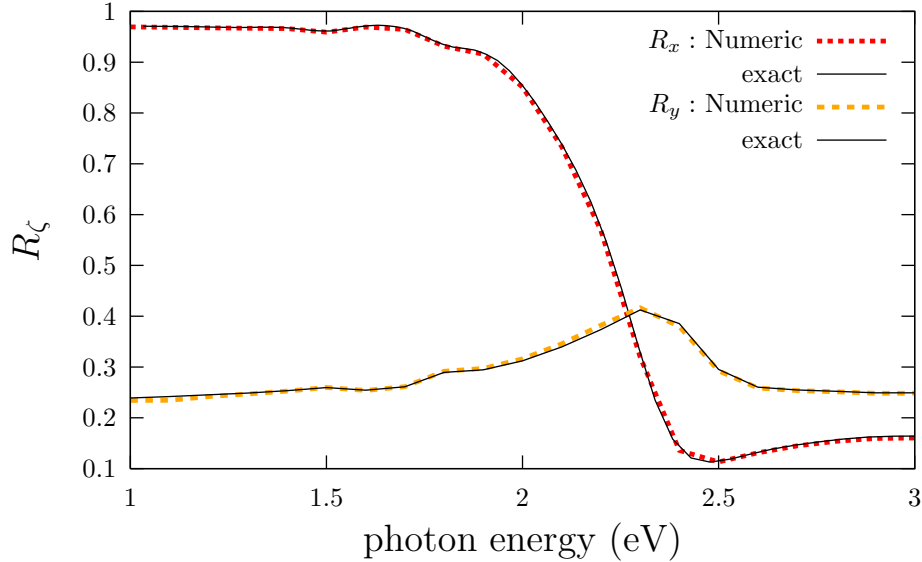}
\caption{(color online) 
Reflectance $R_\zeta$ ($\zeta=x,y$) vs. photon energy   for 
$\xi_x=1$, i.e., for a 1D multi-layer of alternating
slabs of gold ($\epsilon_h=\epsilon(Au)$) and a dielectric ($\epsilon_p=4$), with
$L=40$ nm, and 
$f=0.5$. The slabs are normal to the $y$ direction and the incoming
light propagates along the $z$ direction.
We compare numerical results obtained from Eq.~\eqref{aphiM2} with the
exact non-retarded results obtained through Eq.~\eqref{exM} for $R_x$
and Eq.~\eqref{eyM} or Eq. \eqref{eyMG} (Maxwell-Garnet in 1D) for $R_y$.
}
\label{fig:exactvsnum}
\end{figure}

In Fig.~\ref{fig:exactvsnum} we show $R_\zeta$ vs. the
photon energy $\hbar\omega$ 
as obtained through our
numerical scheme (Eq.~\eqref{aphiM2}), and compare them to the exact
non-retarded results
(Eq.~\eqref{exM} and Eq.~\eqref{eyMG}).
We remark that both
numerical and exact results agree closely.  Actually, in the appendix we show
analytically that
in this case our formalism coincides exactly with Eq. \eqref{exM} and
Eqs. (\ref{eyM}-\ref{eyMG}) and that Eq. \eqref{exM} holds also along the
translational invariant direction of 2D systems.\cite{krokhinPRB02} 

The
system is highly anisotropic ($R_x\ne R_y$) so that the 1D
MG results are only applicable along the $y$ direction ($R_y$).
Notice that the behavior of the system at low
frequencies is metallic for $\zeta=x$, with a very high reflectance,
while it is dielectric-like for $\zeta=y$, as electric the current may
flow unimpeded through the Au layers in the $x$ direction, but it
would be interrupted along the $y$ direction by the dielectric layers.

\begin{figure}
\centering
\includegraphics{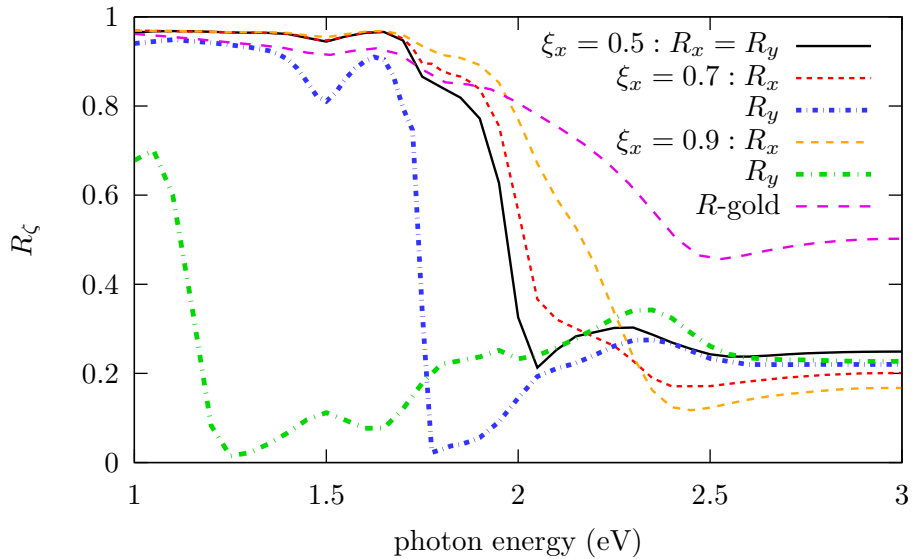}
\caption{(color online) Reflectance $R_x$ (thin lines) and $R_y$
  (thick lines) vs the photon energy,
for a gold host with inclusions of $\epsilon_p=4$ and a fixed
$\xi_y=0.5$, for three different values of $\xi_x$=0.5, 0.7, and 0.9
(see text for details). We also show the reflectivity of gold. }
\label{fig:rvsxi}
\end{figure}

Having checked that our approach coincides with two well-known analytic
limits, we proceed to show results for other choices of $\xi_x$
and $f$. 
In
Fig.~\ref{fig:rvsxi} we show the reflectance $R_\zeta$
for  inclusion with three  rectangular cross sections
with a fixed $\xi_y=0.5$ for several choices of $\xi_x$=0.5, 0.7, and
0.9, with corresponding values of 
$f$=0.25, 0.35, and 0.45. Thus, we  include square and rectangular prisms.
As could be expected,  $R_x=R_y$ for the
square isotropic case $\xi_x=\xi_y$, while for rectangular sections
the reflectance becomes strongly dependent on the polarization $\zeta=x$
or $y$; the anisotropy increases as $\xi_x$ moves away from $\xi_y$. 
As $\epsilon_p$ and $\epsilon_h$ are isotropic, the anisotropy
$\epsilon^M_x\ne\epsilon^M_y$ of the macroscopic response 
arises from  the last term of Eq.~\eqref{aphiM2}. Thus, the source of
the anisotropy is the local-field interaction among the inclusions,
linked to the geometry of the system.

We notice that $R_\zeta$ for $\zeta=x$ polarization, along the elongated
side of the rectangles, is qualitatively similar
to the isotropic case, as well as to that of gold (shown in the
same Fig.~\ref{fig:rvsxi}). To wit, for low 
energies the reflectance is very large, as gold behaves like a Drude
metal and most of the light is reflected. For higher energies 
 and especially  above the
interband-transitions threshold of Gold ($\sim 2.44$ eV), the
reflectance diminishes as gold deviates from the pure Drude-like
behavior and dissipation mechanisms beyond ohmic heating appear. It is
important to note that the surface and bulk plasma  
frequencies ($\sim$ 5, 6 eV) are still higher up in energy  than such
threshold. 

However, we notice an interesting effect for $\zeta=y$  polarization,
along the short side of the rectangles. At some energies $R_y$ deviates
strongly from the isotropic case as $\xi_x$ increases, and shows a
counterintuitive behavior, developing a deep minimum which approaches
zero reflectance for some values of the photon energy.
This may appear surprising, as gold is very reflective in the infrared
region. Nevertheless the geometry of the inclusions changes this
behavior dramatically. It is also interesting to note that above
the interband threshold the anisotropy is drastically 
reduced as $R_x\sim R_y$.

\begin{figure}
\centering
\includegraphics{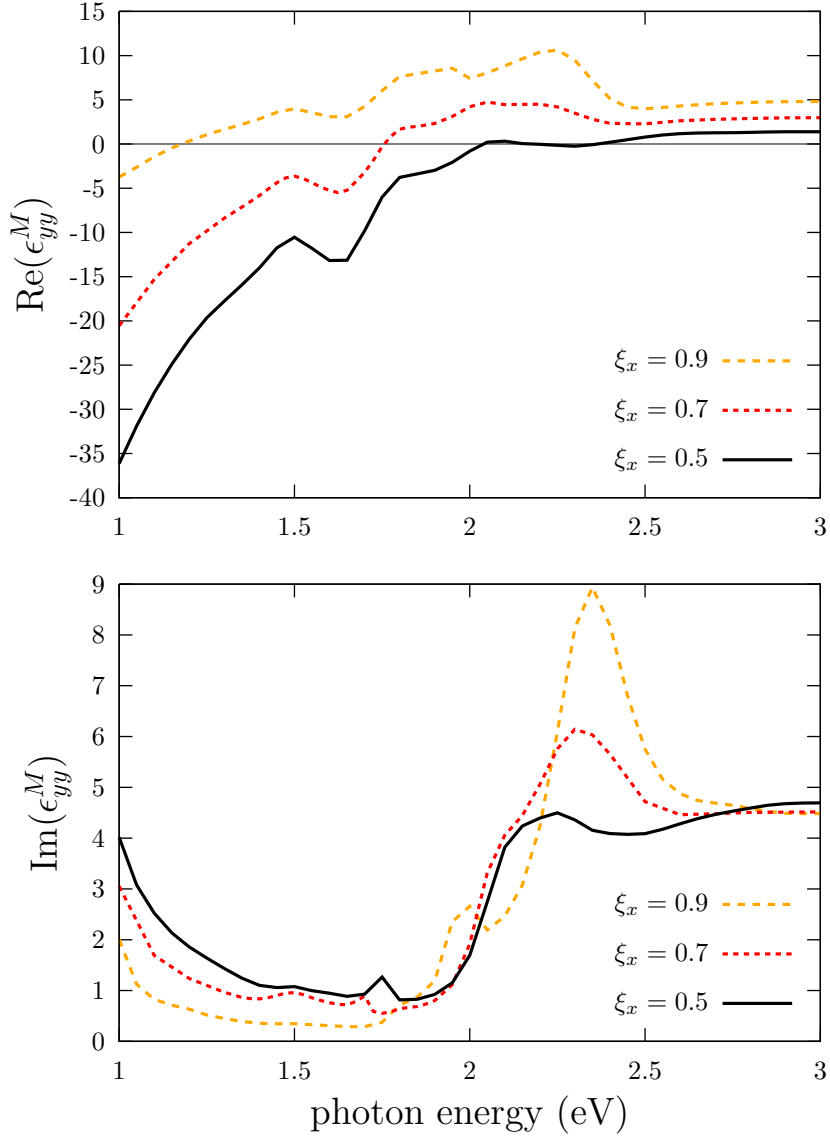}
\caption{(color online)  Real (top panel) and imaginary
  (bottom panel) part of
  $\epsilon^M_{yy}$ vs. photon energy for the same system  as that presented on 
  Fig.~\ref{fig:rvsxi}.
  }
\label{fig:epsvsxi}
\end{figure}

To explain the surprising behavior of the reflectance,  in Fig.~\ref{fig:epsvsxi}
we show the real and imaginary part of the macroscopic response
$\epsilon^M_{yy}$  
for the same system as the one presented in Fig.~\ref{fig:rvsxi}.
For
$\xi_x>\xi_y$ the response along $y$ is 
dielectric like, with a positive Re($\epsilon^M_{yy}$) larger than unity, not
unlike the 1D layered system presented in Fig. \ref{fig:exactvsnum}. Nevertheless,
as the dielectric prisms are completely isolated from each other by
the metallic 
interstices, the Au region percolates and the behavior at low enough
frequencies is metallic, with a negative $\epsilon^M_{yy}$. Thus, there is a
photon energy
where Re($\epsilon^M_{yy}$) crosses through unity. This energy is
red-shifted as $\xi_x$ grows and the metallic behavior disappears
completely at the limit $\xi_x=1$. Thus, for appropriate values of 
$\xi_x$, the crossing may be situated at frequencies too low to excite
interband transitions in Au, but large enough so that ohmic
losses become unimportant. For that frequency at which
$\mbox{Re}(\epsilon_{yy})\approx 1$,
and $\mbox{Im}(\epsilon_{yy})\ll1$ there is a good impedance matching between
vacuum and the material, and thus, there is a small reflectance which
approaches zero.  When
this conditions holds, the transmittance of a finite slab approaches unity.
Our results show that this is the case at $\hbar\omega\approx 1.25$
(1.7) eV for $\xi_x=0.9$ 
($\xi_x=0.7$).

\begin{figure}
\centering
\includegraphics{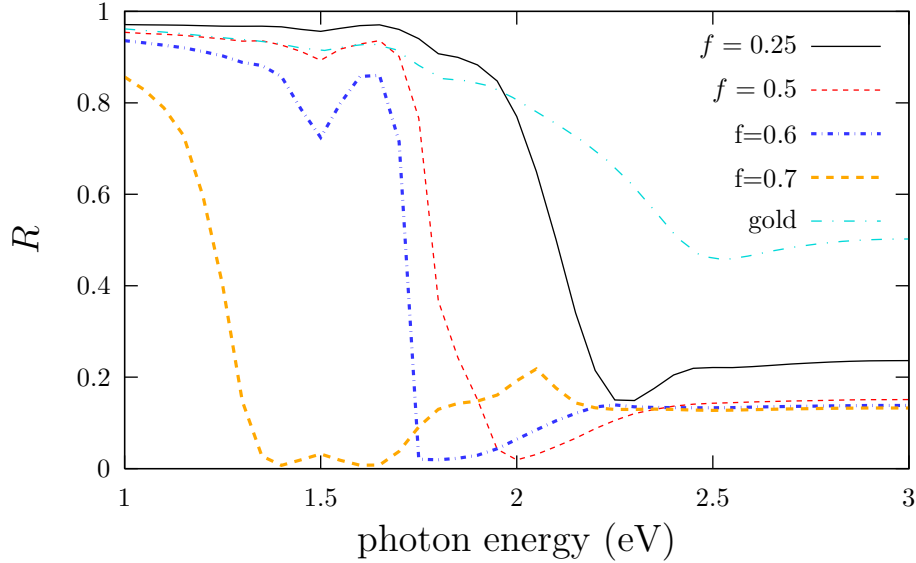}
\caption{(color online) 
Reflectance  $R_x=R_y=R$ 
vs. photon energy for a system made up of circular cylinders with
$\epsilon_p=4$ within an Au matrix
$\epsilon_h=\epsilon(\mbox{Au})$ for four values of the filling fraction
$f=0.25, 0.5, 0.6,$  and 0.7. We also
show the reflectance  of gold for comparison.  
}
\label{fig:circ}
\end{figure}

We explained the different behaviors between the $x$ and $y$ response
of a system of rectangular prisms for different values of $\xi_x$ and
different frequencies in terms of a low-frequency metallic and
high-frequency insulating behavior of the composite, which in turn is 
related to the percolation of the metallic host. We may confirm these
ideas by studying other systems with dielectric inclusions within a
percolating conducting host, such as a square array of dielectric
cylinders within an Au host. 
For a filling fraction $f>\pi/4$ the dielectric cylinders would touch each
other impeding the flow of current between the conducting
regions that would become isolated from each other, and the system
would behave as an insulator. Thus, we study the case $f<\pi/4$ for
which we expect the low frequency behavior to be metallic like with a
transition into a dielectric like behavior at higher frequencies as in
the rectangular case. This system is however isotropic, $R_x=R_y=R$.
To perform the calculation we only require the Fourier coefficients
$S(\mathbf{G})=2fJ_1(Gr_c)/Gr_c$, with $r_c$ the radius of the  inclusions,
and $J_1$ the $J$-Bessel function of order one. 
Fig.~\ref{fig:circ} shows that $R$ indeed attains large values at low
frequencies, corresponding to a metallic behavior, and then attains a
minimum corresponding to the expected transition into a dielectric
behavior.  The transition frequency is red-shifted and becomes broader
and deeper as $f$ increases.

Our examples show that the reflection goes rapidly from almost one to
almost zero at a frequency in the near infrared which may be tuned by
choosing the filling fraction and, in the case of rectangular
inclusions, by changing the aspect ratio. 
A square array of cylinders is isotropic within the $x-y$ plane, so,
in a sense, the array of rectangular prisms is richer, as it
allows us to change the behavior from conducting-like to
insulating-like by simply rotating the polarization.

We remark that the behavior of the reflection discussed above is
induced solely by the geometry of the metamaterial 
and is not simply connected to the structure of the response
functions of the constituent
materials, that is, to  
resonances in $\epsilon_p$ and/or in $\epsilon_h$. Similar resonances, mainly
related to the geometry of the metamaterial, were already predicted
by Khizhnyak back in 1958.\cite{Khizhnyak(1959)}
Our results, clearly show the huge difference that the shape of the inclusions
makes on the optical properties of the
system.\cite{kleinPRL04,gordonPRL04,Garcia(2005)} 

\section{Conclusions}\label{conc}

We have developed a systematic scheme to calculate
the complex frequency dependent 
macroscopic dielectric function for
metamaterials. 
Starting from Maxwell's
equations and employing a long wavelength approximation we have
derived an expression for the macroscopic dielectric 
function $\epsilon^M_{ij}$ 
 that depends on the dielectric functions of the
host $\epsilon_h$ and
particles $\epsilon_p$, and on the geometry of both the unit cell and the
inclusions. The calculation is setup through expansions of the
microscopic fields in plane wave components, and in
general a large number of reciprocal vectors  $\mathbf{G}$ are required to achieve
convergence of the results. We validated our formalism through
convergence tests and through
comparison of our  results to those from previous 
calculations, founding an excellent agreement. 
Then, we calculated macroscopic response and the normal-incidence reflectivity
for systems made up of dielectric rectangular prisms and cylinders 
arranged in a 2D
square lattice  within a gold host. Although the host and the
inclusions are intrinsically isotropic, we
found that, 
if the inclusion is geometrically anisotropic, so is the
macroscopic optical response. For 
rectangular prisms of high aspect ratio
we found a very  anisotropic optical
response, where the infrared reflectance
is almost unity when the field is polarized along the 
long axis, while it can attain values very close to zero when the field is
polarized along the short axis. We explained this
behavior in terms of a transition from a low-frequency conducting
behavior to a high-frequency dielectric behavior for systems not too
far from percolation into the non-conducting phase. The transition may
occur at frequencies in the infrared frequencies for which one would naively
expect very low values for the transmittance. We verified this
explanation through the calculation of the reflectance of a square array of
cylindrical prisms, which shows an isotropic but otherwise similar
behavior as we approach the percolation threshold $f=\pi/4$. Our
formalism may be employed to explore and design of very diverse systems with
a tailored optical response. We hope this work would motivate the
 construction of such systems and their optical
characterization for the  experimental verification of our results.     

\acknowledgments

We acknowledge inspiring discussions with Peter Halevi and Felipe
P\'erez-Rodr\'{\i}guez. This work was partially supported by DGAPA-UNAM
grant IN120909 (WLM), by CONACyT grants 48915-F (BMS) and J49731-F
(BEMZ) and by ANPCyT grant 190-PICTO-UNNE (GPO).

\appendix

\section*{Appendix}

We show that our formalism, as embodied in Eq. \eqref{Phi2},
Eq. \eqref{aphiM2} and Eq. \eqref{W0} are equivalent in the
non-retarded limit to the
analytical results 
\eqref{exM} and Eq. \eqref{eyM} for the case of periodically alternating
isotropic thin flat slabs. We chose the $y$ axis normal to the slabs,
so that the reciprocal vectors $\mathbf{G}=G \hat y$ lie all along $y$. In
this case, both  
$[\mathcal W_0(G,G')]_{ij}$ 
and $[\epsilon_0(G,0)]_{ij}$ are diagonal, so we can
consider separately the cases of 
polarization along the $x$ and the $y$ axes. 

For $x$ polarization we
rewrite Eq. \eqref{Phi2} as
\begin{equation}\label{xx1}
\sum_{G'\ne0}\left(\epsilon_h\delta_{GG'}+\epsilon_{ph} S(G-G')-\frac{G^2}{k_0^2}
\delta_{GG'}\right) [\Phi_0(G',0)]_{xx} = \epsilon_{ph} S(G), 
\end{equation}
whose solution is
\begin{equation}\label{xx0}
  [\Phi_0(G',0)]_{xx} = 0+\mathcal O(k_0^2/G^2).
\end{equation}
Substitution in Eq.
\eqref{aphiM2} yields immediately Eq. \eqref{exM} to order 0 in
the small quantities $k_0/G$ in the non-retarded limit. Notice that
the argument above is valid for any system which has translational
invariance along one or more directions whenever the polarization
direction points along those directions, since Eq. \eqref{xx1} holds
when all the reciprocal vectors 
$\mathbf{G}$ are perpendicular to the polarization direction. In particular,
for systems which have texture only along two dimensions, the
macroscopic dielectric function along the third dimension is simply
the volume average of the microscopic dielectric functions.\cite{krokhinPRB02}  

For $y$ polarization we rewrite Eq. \eqref{Phi2} as
\begin{equation}\label{yy1}
  \sum_{G'\ne0}\left(\epsilon_h\delta_{GG'}+\epsilon_{ph} S(G-G')\right)
      [\Phi_0(G',0)]_{yy} = \epsilon_{ph} S(G),  
\end{equation}
as $G^2-G_y G_y=0$. Although $[\Phi_0(G,0)]_{yy}$ is only defined for
$G\ne 0$, we can extend its definition to $G=0$ by choosing
$[\Phi_0(0,0)]_{zz}\equiv 0$ and extending Eq. \eqref{yy1} to include
the $G=0$ term. For consistency, we have to add an unknown term to its
RHS which only applies to the $G=0$ term, i.e.,
\begin{equation}\label{yy2}
  \sum_{G'}\left(\epsilon_h\delta_{GG'}+\epsilon_{ph} S(G-G')\right)
      [\Phi_0(G',0)]_{yy} = \epsilon_{ph} S(G) + C \delta_{G,0}, 
\end{equation}
where the sum includes now all values of $G'$. Taking the Fourier
transform of Eq. \eqref{yy2} we obtain
\begin{equation}\label{yy3}
  \epsilon_h[\Phi_0(y)]_{yy}+\epsilon_{ph} S(y) [\Phi_0(y)]_{yy}
       = \epsilon_{ph} S(y) + C, 
\end{equation}
which yields
\begin{equation}\label{yy4}
  [\Phi_0(y)]_{yy} = \frac{\epsilon_{ph} S(y) + C}{\epsilon(y)}.
\end{equation}
The constant $C$ must be chosen so that the spatial average of
$[\Phi_0(y)]_{yy}$ vanishes, 
\begin{equation}\label{yy5}
0=[\Phi_0(G=0,0)]_{yy} = \frac{1-f}{\epsilon_h} C + \frac{f}{\epsilon_p}(\epsilon_{ph}+C),
\end{equation}
vanishes. Substituting the result in \eqref{yy4} we obtain
\begin{equation}\label{yy6}
  [\Phi_0(y)]_{yy} = \frac{\epsilon_{ph}}{\epsilon(y)} \left(S(y)-\frac{f
    \epsilon_h}{\epsilon_h+\epsilon_{ph}(1-f)}\right). 
\end{equation}
Now we extend the sum in Eq. \eqref{aphiM2} to include the $G=0$
contribution, allowing us to employ the convolution theorem to obtain
\begin{equation}\label{yy7}
  \sum_G S(-G) [\Phi_0(G,0)]_{yy} =
  \frac{1}{L_y}\int dy\, S(y)
       [\Phi_0(y)]_{yy} 
       =f \frac{\epsilon_{ph}}{\epsilon_p} \left(1-\frac{f
         \epsilon_h}{\epsilon_h+\epsilon_{ph}(1-f)}\right), 
\end{equation}
which we substitute into Eq. \eqref{aphiM2} to finally obtain Eq. \eqref{eyMG}.

\bibliography{referencias}
\end{document}